\documentclass[11pt,twoside,a4paper]{article}
\usepackage{amsmath,amssymb}
\usepackage[dvips]{graphicx}
\usepackage[twoline]{mya4}

\newcommand{\scrF}{\ensuremath{\mathcal{F}}}
\newcommand{\etaG}{\ensuremath{\eta_G}}
\newcommand{\MS}{\ensuremath{M_{\text{S}}}}
\newcommand{\MPl}{\ensuremath{M_{\text{Pl}}}}
\newcommand{\MEW}{\ensuremath{M_{\text{EW}}}}
\newcommand{\mm}{\ensuremath  {\ \text{mm}}}
\newcommand{\nm}{\ensuremath  {\ \text{nm}}}
\newcommand{\TeV}{\ensuremath {\text{\ Te\kern -0.1em V}}}
\newcommand{\GeV}{\ensuremath {\text{\ Ge\kern -0.1em V}}}
\newcommand{\MeV}{\ensuremath {\text{\ Me\kern -0.1em V}}}
\newcommand{\keV}{\ensuremath {\text{\ ke\kern -0.1em V}}}
\newcommand{\MET}{\mbox{\ensuremath{\,\slash\kern-.7em E_{\text{T}}}}}
\newcommand{\id}{\ensuremath{\text{d}}}
\newcommand{\ppbar}{\ensuremath{\text{p}\bar{\text{p}}}}
\newcommand{\pp}{\ensuremath{\text{pp}}}
\newcommand{\invpb}{\ensuremath{\ \text{pb}^{-1}}}
\newcommand{\invfb}{\ensuremath{\ \text{fb}^{-1}}}
\newcommand{\Zob}{\ensuremath{\mathrm{Z}}}
\newcommand{\dzero}{D\O}

\begin{document}

\begin{flushright}
{\large MAN/HEP/2003/7}
\end{flushright}

\vskip 2em
\begin{center}
{\LARGE
Search for Extra Dimensions at Hadron Colliders}
\vskip 1.5em
{\large
Michiel Sanders (for the CDF and \dzero{} collaborations) 
\footnote{Presented at the
International Europhysics Conference on High Energy Physics, Aachen,
Germany, July 2003.} \\[.5em]
University of Manchester, United Kingdom}
\vskip 1em
{\large
14th October 2003}
\vskip 1.5em
\end{center}

\begin{abstract}
To explain the large difference between the Planck scale and the
electroweak scale, models in which gravity propagates in more than
four dimensions have recently been proposed. In this paper, results
from searches for extra dimensions at the Tevatron \ppbar{} collider
are presented. Limits are set on the higher dimensional Planck scale,
and expectations for the LHC \pp{} collider are given.
\end{abstract}

\section{Introduction}
\label{intro}

Precision tests have shown a remarkable
agreement between experimental data and predictions from the Standard
Model, especially in the
electroweak sector \cite{PippaWells}. However, the Standard Model does
not explain the huge value of the Planck scale (\MPl{}) with respect
to the electroweak scale (\MEW{}): $\MPl/\MEW \sim 10^{16}$.
It has recently been pointed out that this hierarchy problem can be
solved if gravity propagates in more dimensions than the Standard
Model particles.

Arkani-Hamed, Dimopoulos and Dvali (ADD) proposed a model \cite{ADD} in which
gravity propagates freely in $n$ extra, compact
spatial dimensions. In their model, the electroweak scale
is the only fundamental scale, and the apparent large Planck scale
follows directly from the existence of extra dimensions. 
From a generalization of
the gravitational potential in $4+n$ dimensions, they find for the
radius $R$ of these extra dimensions: 
$R \sim \frac{1}{\MS}( \MPl/\MS )^{2/n}$, 
where \MS{} is the fundamental scale in $4+n$ dimensions, and \MPl{}
is the apparent 4-dimensional Planck scale. Taking $\MS \simeq \MEW$,
$R\simeq 1\mm$ for $n=2$ and $R \lesssim 1\nm$ for $n \geq 3$. It is
remarkable that from direct gravitational measurements, only 
$n \leq 2$ has been excluded at present \cite{SalvatoreMele}. 
Note that  in the ADD approach,
the size of the extra dimensions is ``large'' with respect to
$1/\MEW$. 

Randall and Sundrum (RS) propose a 5-dimensional model \cite{RS} in
which the 4-dimensional metric is multiplied by an exponential
``warp'' factor $e^{-2 k r_c \phi}$, where $k$ is the 5-dimensional
Planck scale. The fifth dimension appears as a finite interval 
$0 \leq \phi \leq \pi$ whose size is given by $r_c$. Even if $r_c$ is
small, a large hierarchy can be generated by this exponential factor.
In other words, the size of the extra dimension in the RS model 
is relatively small
compared to the size of the extra dimensions in the ADD approach.

In both models, effects of the extra dimensions could become visible
in collider experiments at the\TeV{} scale.

\subsection{Experimental Signatures}

In hadron-hadron collisions, real gravitons can be produced in
association with a jet or a photon \cite{GRW}. The graviton propagates
in the extra dimensions and is therefore invisible in a detector. It
will manifest itself as missing transverse energy (\MET{}).

Virtual graviton exchange will modify the cross section for di-lepton
and di-photon production. The exact experimental observation is
different for the two models.
In the RS model, the size of the extra dimension is small and thus the
separation between the 
Kaluza-Klein modes (quantized energy levels in the finite extra dimension)
of the graviton is large. 
Therefore, the graviton will be produced resonantly. However,
in the ADD model the separation between the graviton Kaluza-Klein
modes is small (as small as $2\keV$ for $n=4$ \cite{GRW}) due to the
size of the extra dimensions. This will lead to an overall modification
of the cross section at high di-lepton or di-photon invariant
mass. 

The cross section for di-lepton or di-photon production in the
presence of extra dimensions can be written as
\begin{equation}
\frac{\id^2\sigma}{\id M \id \cos\theta^*} = 
     f_{\text{SM}} + f_{\text{int}} \etaG + f_{\text{KK}} \etaG^2, 
\label{eq:2d-crosssection}
\end{equation}
with $\etaG = \scrF / M_{\text{S}}^4$. The parameter $M$ is the
invariant mass of
the two particles and $\theta^*$ is the scattering angle in the
centre-of-mass frame. The term $f_{\text{SM}}$ describes the Standard
Model cross section, $f_{\text{int}}$
the interference of the graviton-induced amplitude with that of
the Standard Model
processes, and $f_{\text{KK}}$ describes the pure graviton exchange term.
Different formalisms exist for the dimensionless parameter \scrF{}:
\begin{eqnarray}
\scrF &=& 1,  \quad (\text{GRW \cite{GRW}}); \label{eq:GRW} \\
\scrF &=& \left\{\begin{array}{ll}
	\log (\frac{\MS^2}{M^2})&\, n=2 \\
	\frac{2}{n-2} &\, n > 2
	\end{array}   \right.,  \quad (\text{HLZ \cite{HLZ}}); \label{eq:HLZ} \\
\scrF &=& \frac{2\lambda}{\pi} = \pm\frac{2}{\pi},  
       \quad (\text{Hewett \cite{Hewett}}). \label{eq:Hewett}
\end{eqnarray}

In the remainder of this paper, experimental searches for 
graviton emission and
exchange at the Tevatron \ppbar{} collider will be presented. During
Run~I, the \ppbar{} centre-of-mass energy was $1.8\TeV$ whereas during the
ongoing Run~II, it has been increased to $1.96\TeV$, with higher
instantaneous and integrated luminosities. In addition, a 
brief overview of prospects 
at the LHC \pp{} collider ($\sqrt{s} = 14\TeV$) is
given. This is discussed in more detail in \cite{LaurentVacavant}.

\section{Experimental Results}

\subsection{Graviton Emission in the ADD Model}

\begin{figure}
\begin{center}
\includegraphics[width=.5\textwidth,angle=270]{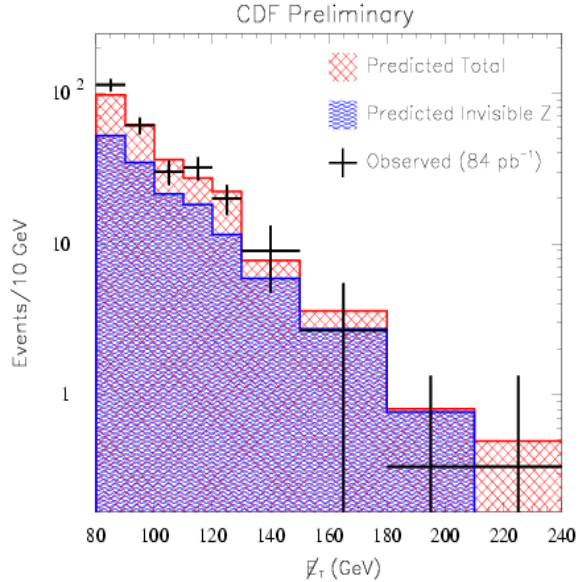}
\end{center}
\caption{Distribution of \MET{} in jets~+~\MET{} events in CDF Run~I data.}
\label{fig:CDFjets}
\end{figure}

The CDF collaboration has searched for graviton emission in
the jets~+~\MET{} channel, based on an
integrated luminosity of $84\invpb$ collected during Run~I. The main
background in this channel is \Zob{}~+~jets production, where the \Zob{}
boson decays invisibly in a neutrino-antineutrino pair. As can be seen
from the \MET{} distribution in Fig.~\ref{fig:CDFjets}, the data agree
well with  expectation in the absence of extra dimensions.

This leads to lower limits on \MS{} of
1\TeV{}, 770\GeV{}, 710\GeV{} for $n=2,\,
4,\, 6$, respectively\footnote{All limits are given at 95\% confidence
level.}. 
These results are very similar
to the previously established \dzero{} limits in this channel \cite{D0jets}.
A similar search by the CDF collaboration in the $\gamma$~+~\MET{} channel
gives no excess above the  expectation from the Standard Model 
and thus lower limits on
\MS{} of 549\GeV{}, 581\GeV{}, 602\GeV{} for $n=4,\,
6,\, 8$, respectively.  The sensitivity in the jets~+~\MET{} sample is
higher due to the larger cross section. 

At the LHC, with $100\invfb$ of integrated luminosity, discovery of
extra dimensions in the jets$\,+\,\MET$ channel is possible up to high
mass scales. As shown by the ATLAS collaboration
\cite{LaurentVacavant}, 
significant signals
can be found up to values of \MS{} of 
9.1\TeV{}, 7.0\TeV{}, 6.0\TeV{} for $n=2,\, 3,\, 4$.

\subsection{Graviton Exchange in the ADD Model}

Both the CDF and \dzero{} collaborations have searched for virtual
graviton exchange in the ADD model. CDF analyzed the invariant mass
spectrum of photon pairs from an integrated luminosity of
approximately $95\invpb$, collected during Run~I. The selected sample
agrees well with the expectation from Standard Model induced di-photon
production and from misidentified di-photon events. A similar agreement was
found in a di-electron sample, leading to lower limits on
\MS{} in the Hewett convention (see Eq.~(\ref{eq:Hewett})) given in  
Table~\ref{tab:CDFdiEM-ADD}.
\begin{table}
\begin{center}
\begin{tabular}{ccc}
\hline\noalign{\smallskip}
                & $\lambda=-1$ & $\lambda=+1$ \\ 
\noalign{\smallskip}\hline\noalign{\smallskip}
ee              & 826    & 808       \\
$\gamma\gamma$  & 899    & 797       \\
di-EM           & 939    & 853       \\
\noalign{\smallskip}\hline
\end{tabular}
\end{center}
\caption{Lower limits on \MS{} in\GeV{} (Hewett convention) in the di-electron,
di-photon and combined samples from CDF Run~I.}
\label{tab:CDFdiEM-ADD}
\end{table}

The \dzero{} collaboration presented results on di-muon events in
30\invpb{} of integrated luminosity in Run~II. Background to graviton
exchange in this decay mode is mainly caused by Standard Model di-muon
production, estimated from a Monte Carlo simulation, and by heavy
quark decay, estimated from the data itself. Similarly, di-photon and
di-electron events were studied in 120\invpb{} of integrated
luminosity in Run~II. 
Background from
Standard Model di-photon and di-electron production was estimated from
Monte Carlo simulations, and background from misidentified QCD events
was derived from the data sample. In both channels, a good agreement 
between data and background estimates was observed in distributions
of both the invariant mass and the scattering angle. 

\begin{figure}
\begin{center}
\includegraphics[width=.55\textwidth]{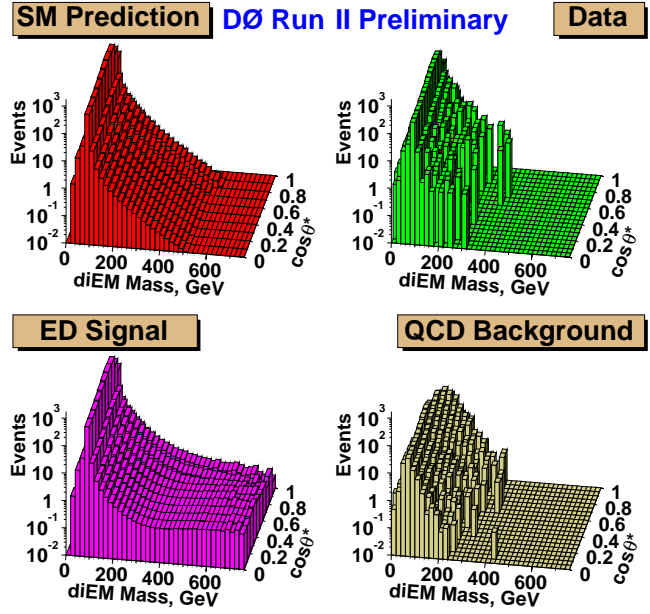}
\end{center}
\caption{Distributions of invariant mass of electron and photon pairs
versus scattering angle in \dzero{} Run~II data for (top-left)
Standard Model expectation, (top-right) data, (bottom-left) Standard
Model plus an extra dimensions signal, and (bottom-right) QCD
background.}
\label{fig:D0diEM-ADD}
\end{figure}

A 10-15\% higher sensitivity to graviton exchange is expected from
fitting the data in the $M$ versus $\cos\theta^*$ plane \cite{LandsbergCheung} 
instead of in $M$ only. This is illustrated in
Fig.~\ref{fig:D0diEM-ADD}. In the di-muon sample, such a
two-dimensional fit (see Eq.~(\ref{eq:2d-crosssection})) 
yields $\etaG = 0.02 \pm 1.35 \TeV^{-4}$  and in the combined di-photon,
di-electron sample $\etaG = 0.0 \pm 0.15 \TeV^{-4}$.
This translates into lower limits on \MS{} given in 
Table~\ref{tab:D0diEMdiMu-ADD}.

\begin{table}
\begin{center}
\begin{tabular}{lcccc}
\hline\noalign{\smallskip}
          & GRW  & \multicolumn{2}{c}{HLZ} & Hewett       \\ 
          &      & $n=2$ & $n=7$           & $\lambda=+1$ \\
\noalign{\smallskip}\hline\noalign{\smallskip}
di-EM     & 1.28 & 1.42  & 1.01            & 1.14         \\         
$\mu\mu$  & 0.79 & 0.68  & 0.63            & 0.71         \\
\noalign{\smallskip}\hline
\end{tabular}
\end{center}
\caption{Lower limits on \MS{} in\TeV{} in the combined di-electron
and di-photon sample and in the di-muon sample from \dzero{} Run~II data.}
\label{tab:D0diEMdiMu-ADD}
\end{table}

\subsection{Graviton Resonance Production in the RS Model}

\begin{figure}
\begin{center}
\includegraphics[width=.5\textwidth]{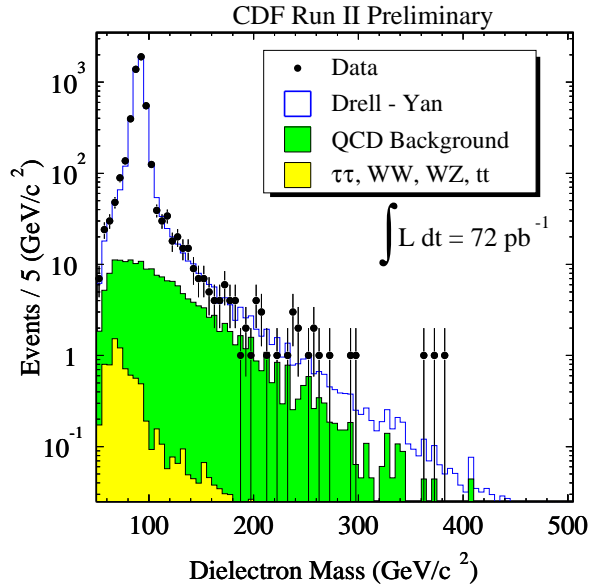}
\end{center}
\caption{Invariant mass distribution of electron pairs in CDF Run~II data.}
\label{fig:CDFdiEl-RS}
\end{figure}

The CDF collaboration has performed a search for graviton resonance
production in the di-muon and di-electron decay channels, using an
integrated luminosity of 72\invpb{} collected during Run~II. The invariant mass
distribution found in this data in both channels agrees well with the
expectation from the Standard Model, as shown in
Fig.~\ref{fig:CDFdiEl-RS} for the di-electron channel.
This leads to lower limits on the RS graviton, given in 
Table~\ref{tab:CDF-RS}.
The ATLAS collaboration expects a sensitivity up to a mass of 2\TeV{}
in 100\invfb{} of integrated luminosity at the LHC \cite{LaurentVacavant}.

\begin{table}
\begin{center}
\begin{tabular}{ccc}
\hline\noalign{\smallskip}
ee  & $\mu\mu$ & combined \\
\noalign{\smallskip}\hline\noalign{\smallskip}
535 & 370      & 550 \\
\noalign{\smallskip}\hline
\end{tabular}
\end{center}
\caption{Lower limits on RS graviton mass in\GeV{} for $k/\MPl = 0.1$, 
in the di-electron and di-muon channels, and combined, from CDF Run~II data.}
\label{tab:CDF-RS}
\end{table}

\section{Conclusion and Outlook}

To explain the large difference between the Planck scale and the
electroweak scale, models have been proposed in which gravity
propagates in more than four dimensions. The effects of these higher
dimensional models were searched for at the Tevatron, and were not
found, yet. Lower limits on the higher dimensional Planck scale were
set, reaching over 1\TeV{} in current Tevatron Run~II data. 
The
combined di-photon and di-electron sample of \dzero{}
provides the most stringent limits on the ADD model.
In the near future, more data from
Run~II at the Tevatron will become available, increasing the
sensitivity to extra dimensions. After 2007, the experiments at the
LHC will take over and explore and possibly discover extra dimensions
up to multi-$\!\!$\TeV{} scales.


\begin{thebibliography}{}



\bibitem{PippaWells}
Pippa Wells, this conference.
%
\bibitem{ADD}
N.~Arkani-Hamed, S.~Dimopoulos, G.~Dvali, 
Phys.~Lett.~B \textbf{429}, (1998) 263.
%
\bibitem{SalvatoreMele}
Salvatore Mele, this conference.
%
\bibitem{RS}
L.~Randall, R.~Sundrum, 
Phys.~Rev.~Lett. \textbf{83}, (1999) 3370.
%
\bibitem{GRW}
G.F.~Giudice, R.~Rattazzi, J.D.~Wells, 
Nucl.~Phys.~B \textbf{544}, (1999) 3.
%
\bibitem{HLZ}
T.~Han, J.~Lykken, R.~Zhang, 
Phys.~Rev.~D \textbf{59}, (1999) 105006.
%
\bibitem{Hewett}
J.~Hewett, Phys.~Rev.~Lett. \textbf{82}, (1999) 4765.
%
\bibitem{LaurentVacavant}
Laurent Vacavant, this conference.
%
\bibitem{LandsbergCheung}
K.~Cheung, G.~Landsberg, Phys.~Rev.~D \textbf{62}, (2000) 076003
%
\bibitem{D0jets}
\dzero{} collaboration, V.M.~Abazov et al., 
Phys.~Rev.~Lett. \textbf{90}, (2003) 251802.


\end{thebibliography}
\end{document}